# The Density PDFs of Supersonic Random Flows

By ÅKE NORDLUND[1,2] AND PAOLO PADOAN[3]

[1]Theoretical Astrophysics Center, Juliane Maries Vej 30, 2100 Copenhagen Ø, Denmark

[2]Astronomical Observatory / NBIfAFG, Juliane Maries Vej 30, 2100 Copenhagen Ø, Denmark

[3]Instituto Nacional de Astrofísica, Óptica y Electrónica, Apartado Postal 216, 72000 Puebla, México

The question of the shape of the density PDF for supersonic turbulence is addressed, using both analytical and numerical methods. For isothermal supersonic turbulence, the PDF is Log-Normal, with a width that scales approximately linearly with the Mach number. For a polytropic equation of state, with an effective gamma smaller than one, the PDF becomes skewed and becomes reminiscent of (but not identical to) a power law on the high density side.

## 1. Introduction

The Probability Density Function of mass density is an important statistical property of the ISM that relates, for example, to gravitational collapse and star formation. Log-Normal PDFs have been discussed occasionally in both the cosmological and interstellar contexts (Hubble, 1934; Peebles, 1980; Ostriker, 1984; Zinnecker, 1984; Coles & Jones, 1991). Vázquez-Semadeni (1994) noticed that the density PDFs in his 2-D numerical simulations of turbulence were consistent with a Log-Normal, and discussed possible reasons for the lognormality. Padoan et al. (1997) showed that the standard deviation of the Log-Normal PDFs in their isothermal 3-D simulations was approximately equal to half the rms Mach number. Scalo et al. (1998) raised the questions of how a polytropic equation of state, and more generally a realistic ISM cooling function, might influence the PDF.

In this contribution we investigate the question of the shape of the PDF for isothermal and polytropic equations of state, using analytical methods and by looking at results from 3-D simulations of supersonic turbulence. Space does not allow the inclusion of all illustrations that were shown at the meeting, but these are available on the World Wide Web†.

It may be prudent to remind ourselves that real ISM turbulence is neither isothermal, nor polytropic. Three important factors: 1) Equation of state / energy equation: The real ISM has a local temperature that is not a simple function of density but results from the evolution of the thermal energy. 2) Magnetic fields: The effective gamma of the magnetic field is two for compression across the field, and zero for compression along the field. What might the effect on the PDF be? 3) Gravity: Gravity takes over control over the most dense regions. This might be expected to influence the dense side of the PDF significantly.

## 2. Theory

In Nordlund & Padoan (1998) we give a formal proof for the Log-Normality of the mass density PDF for isothermal supersonic turbulence. The proof rests on an exact result for a general stationary process, given by Pope & Ching (1993). They prove that the PDF $P(x)$ may be expressed as

$$P(x) = \frac{C_2}{q(x)} \exp\left(\int_0^x \frac{r(x')}{q(x')} dx'\right), \tag{2.1}$$

where

$$q(x) = \langle \dot{X}^2 | x \rangle / \langle \dot{X}^2 \rangle, \tag{2.2}$$

and

$$r(x) = \langle \ddot{X} | x \rangle / \langle \dot{X}^2 \rangle. \tag{2.3}$$

† URL http://www.astro.ku.dk/~aake/talks/Turb98

Ê $X(t)$ is a stationary, standardized random process, and $\langle Y | x \rangle$ denotes the conditional expectation value obtained by sampling the property $Y$ when $X(t)$ is in the neighborhood of x.

To prove that $P(x)$ is Log-Normal it is thus sufficient to show that $q(x)$ is a constant and that $r(x)$ is linear. These statistical properties indeed follow from the dynamic equations under isothermal conditions, where the log pressure is equal to the log density plus a constant. Under such conditions both the continuity equation and the equation of motion have the property that they do not depend on the mean density; only gradients of the log density enter, as may be seen by writing the continuity equation and the equation of motion in the forms

$$\frac{\partial \ln \rho}{\partial t} = -\mathbf{u} \cdot \nabla \ln \rho - \nabla \cdot \mathbf{u}, \tag{2.4}$$

and

$$\frac{\partial \mathbf{u}}{\partial t} = -\mathbf{u} \cdot \nabla \mathbf{u} - \frac{p}{\rho} \nabla (\ln \rho + \ln \frac{p}{\rho}) + F. \tag{2.5}$$

Given the invariance with respect to mean density, an external force $F$ (assumed independent of the mean density) imposes the same "event" structure on initial conditions that only differ by a constant in $\ln \rho$. $q(x)$ measures the mean rate of change of the density during events and, because the dynamics does not depend on the mean density, $q(x)$ is independent of the mean density level $x$.

To show that $r(x)$ is linear, subdivide a large ensemble into sub-ensembles of varying mean density; they all have the same internal statistics. Form the grand average by summing over subensembles. Each subensemble contains statistically identical events (e.g. shocks), only at different mean densities. $r(x)$ samples the PDF in such a way that the first order result is the slope of the log PDF, the second order result is a constant times the curvature, and a third order term is only present if the PDF is not Log-Normal.

Thus, if and only if the dynamics is independent of the mean density, a Log-Normal PDF is the only consistent one. Conversely, if the dynamics depends on the mean density, then the PDF cannot be Log-Normal.

## 3. PDFs from 3-D experiments

PDFs from experiments with isothermal supersonic turbulence (Fig. 1) agree with the theoretically predicted Log-Normal PDFs. In 3-D experiments with solenoidal forcing, the linear standard deviation is, to within the statistical uncertainty, equal to half the r.m.s. Mach number (note that the full drawn curves in Fig. 1 are *not* fits, but are constructed by setting the linear standard deviation equal to half the r.m.s. Mach number, measured over the same time interval as used for sampling the PDF). Deviations in the PDF wings are due to poor statistics (in particular at low density), and limited numerical resolution (in particular at high density).

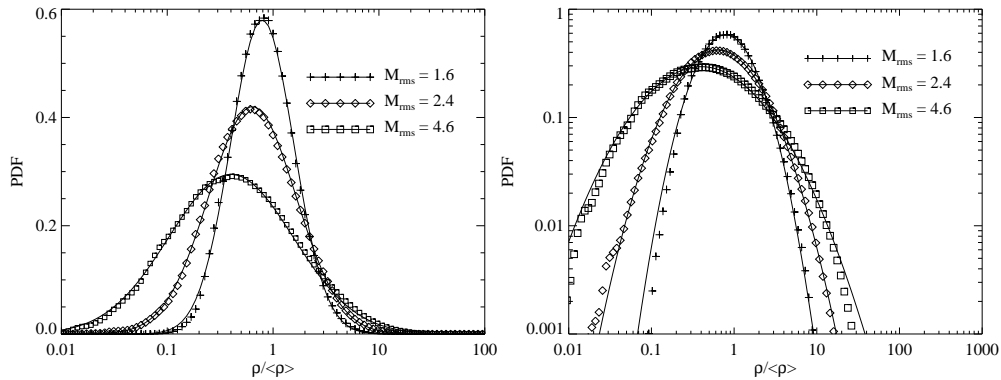

FIGURE 1. Mass density PDFs measured in numerical 3-D experiments of isothermal, supersonic turbulence driven by a solenoidal random force. The PDFs are shown on a lin-log scale (a) and on a log-log scale (b). The symbols show the experimental results. The full drawn curves show Log-Normals with standard deviations equal to half the rms Mach number measured in the experiments.

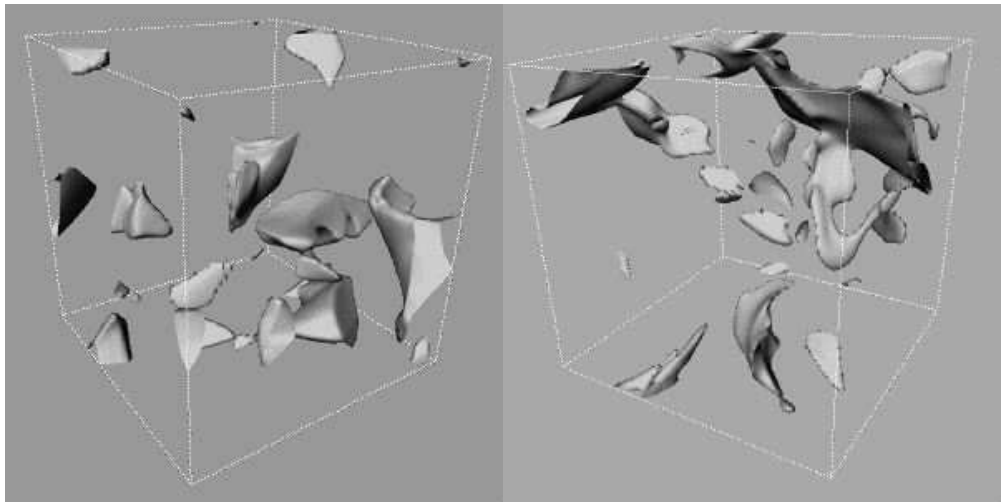

FIGURE 2. Isodensity surfaces in the low (left) and high (right) density wing of the PDF, for a snap shot from one of the numerical experiments. The density levels at which the surfaces are shown are symmetric with respect to the PDF, and correspond to a level about a factor of $10^3$ below the maximum of the PDF.

It is remarkable that, even though the Log-Normal PDF is perfectly symmetrical around its maximum, isodensity surfaces in the two wings of the PDF correspond to very different structures (Fig. 2). High density is created by interaction of 3-D shocks. Structures are sheet fragments and their filamentary and knotty intersections. Low density is created by interaction of 3-D expansion waves. Structures are irregular voids.

It is also remarkable that the high density wing of the Log-Normal is established very early—soon after the first shock interactions. The first shock interactions occur after about 0.2 to 0.3 dynamical times, and after this time the right hand side of the PDF is already well established. For any particular snap shot the PDF contains structure, relative to a perfect Log-Normal, but the right hand side of these early PDFs do not deviate significantly more than PDFs from later times. Features in the PDF progress from high density to low density—presumably these are individual expansion waves.

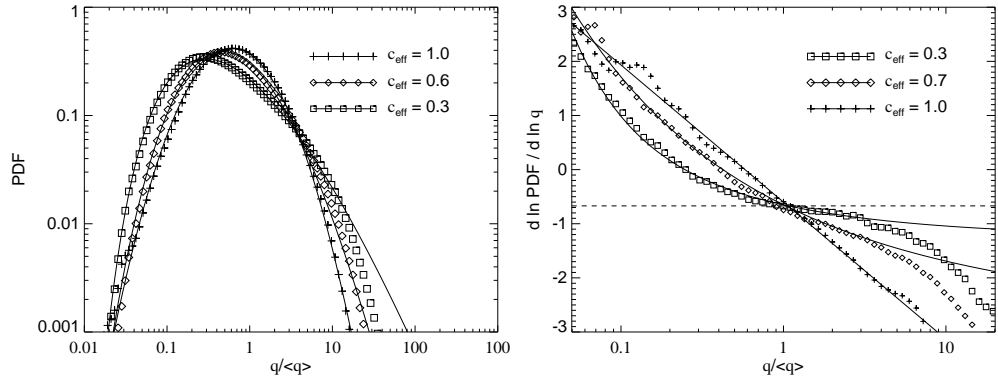

FIGURE 3. PDFs from 3-D experiments with driven supersonic turbulence and a polytropic equation of state. Panel a) shows the PDFs, while panel b) shows their slopes. In both panels, the full drawn curves show analytical fits (cf. discussion in the text). The drop below the analytic fits at high densities is due to limited numerical resolution.

## 4. PDFs for polytropic equations of state

Scalo et al. (1998) pointed out that the PDF is only Log-Normal for isothermal conditions. They argue that the PDF develops a power law wing when the effective gamma is not equal to unity. Their figures 10 and 11 show the PDF from 1-D experiments with effective gammas equal to 1.0 and 0.3, and with varying Mach numbers (note that the Mach numbers given in their paper are not rms Mach numbers, but refer to a non-dimensional parameter in their equations). It is indeed clear from the proof of the Log-Normality of the PDF for isothermal flows that the PDF cannot be exactly Log-Normal for non-isothermal flows, since the dynamic equations are then no longer independent of the mean density. But does the PDF become a power law for a polytropic equation of state?

Driven supersonic 3-D turbulence experiments with effective gamma different from unity produce skewed PDFs (Fig. 3). In the case with an effective gamma less than unity, the dense gas is colder than average, and hence a certain velocity distribution corresponds to higher Mach numbers. A higher Mach number corresponds to a broader PDF, and hence for $\gamma_{\rm eff} < 1$ the PDF has a more extended high density wing than in the isothermal case. The polytropic PDFs are reminiscent of power laws over a limited range of densities; Scalo et al. (1998) fitted the $\gamma_{\rm eff} = 0.3$ PDF with a power law over a range of about one order of magnitude in density, but made no attempt to model the PDF over a larger range of densities.

It is possible to fit the numerical PDFs with an analytical expression, assuming that the logarithmic slope of the PDF is a function of the formal temperature $T \sim \rho^{(\gamma_{\rm eff}-1)}$. Assuming that the slope scales with $T^p$, one can fit the slopes for $p = 5/3$ (cf. Fig. 3). The resulting PDFs are neither Log-Normal, nor power laws.

As is obvious from first principles, and as illustrated by Fig. 3, the family of polytropic PDFs depends in a continuous manner on $\gamma_{\rm eff}$, changing gradually from the symmetric Log-Normal for $\gamma_{\rm eff} = 1$ to more skewed forms for effective gammas that differ from unity. But even the $\gamma_{\rm eff} = 0.3$ PDF differs by less than a factor of two from a Log-Normal, except far out in the wings. It has a "most common density" that is about a factor of two smaller than the one expected for $\gamma_{\rm eff} = 1$.

The continuous change of shape with $\gamma_{\rm eff}$ means that the PDFs cannot be power laws for any finite density (and non-zero $\gamma_{\rm eff}$). The PDFs for $\gamma_{\rm eff} \neq 1$ do have power law asymptotes, but only because the temperature formally goes to zero at one infinity. In reality, the temperature of a 10 K cloud is not expected to fall by more than a factor 3–4 before internal shielding and / or heating become significant.

## 5. The influence of magnetic fields

The Log-Normal like shapes of the PDFs are not noticeably influenced by the presence of weak magnetic fields; as long as the mean magnetic energy remains small, the density PDFs are practically unaffected. For magnetic energies approaching, but still smaller than the mean kinetic energy, the PDFs remain Log-Normal in shape, but with reduced standard deviations, corresponding roughly to the suppression of compressive motions in two out of the three spatial dimensions. For magnetic fields approaching and exceeding equipartition, the PDFs first become truncated at small densities, and then loose their Log-Normal like shape altogether. This has important diagnostic implications, for example for extinction statistics (cf. Padoan & Nordlund, these proceedings).

## 6. Conclusions

Based on the results of Pope & Ching (1993) one may show that the density PDF for supersonic, isothermal turbulence is exactly Log-Normal. Numerical experiments show that, with a solenoidal forcing, and in three dimensions, the standard deviation of the density is equal to half the rms Mach number. Note that, for compressional forcing at low Mach numbers (leading to an ensemble of sound waves), the standard deviation is expected to be equal to the the rms Mach number itself.

The Log-Normal property is robust, in the sense that the PDFs of polytropic supersonic turbulence are not far from Log-Normal. They are skewed, because (for $\gamma_{\text{eff}} < 1$) the dense gas is colder and hence has a larger than average Mach number. They are not power laws for densities of interest, but formally approach power laws as the temperature goes to zero. The presence of a magnetic field changes the standard deviation of the PDF, and modifies the shape somewhat, but a Log-Normal PDF is still a good approximation, as long as the turbulence is super-Alfvénic. Gravity changes the situation qualitatively, in that there may no longer exist stationary turbulence solutions. Empirically, the PDFs tend to approach power laws in the dense wing. The region around the maximum of the PDF is still well described by a Log-Normal.

The near Log-Normal PDF is one of several generic properties of supersonic turbulence—there is thus good hope for rapid progress in our understanding of this subject, which is again important for a better understanding of Interstellar Turbulence.